\documentstyle[11pt,bookman,epsf]{article}
\def\singlespace {\smallskipamount=3.75pt plus1pt minus1pt
                  \medskipamount=7.5pt plus2pt minus2pt
                  \bigskipamount=15pt plus4pt minus4pt
                  \normalbaselineskip=15pt plus0pt minus0pt
                  \normallineskip=1pt
                  \normallineskiplimit=0pt
                  \jot=3.75pt
                  {\def\smallskip {\vskip\smallskipamount}}
                  {\def\medskip   {\vskip\medskipamount}}
                  {\def\bigskip   {\vskip\bigskipamount}}
                  {\setbox\strutbox=\hbox{\vrule
                    height10.5pt depth4.5pt width 0pt}}
                  \parskip 4.5pt
                  \normalbaselines}		  
\def\be{\begin{equation}}
\def\lan{\left\langle}
\def\ran{\right\rangle}
\def\ee{\end{equation}}
\def\barr{\begin{array}}
\def\earr{\end{array}}

\def\nn8{\nonumber\\[10pt]}
\def\l{\left}
\def\r{\right}
\def\dis{\displaystyle}
\def\ed{\end{document}}

\def\cn{{\cal N}}
\def\b0p{\mbox{\boldmath{$0^+$}}}
\begin{document}
\singlespace
\begin{center}
{\bf Random interactions in nuclei and extension of $\b0p$ dominance 
in ground states to irreps of group symmetries}
\vskip 0.25cm
V.K.B. Kota\footnote{Invited talk in the International symposium on `Nuclear
Physics' held at Dalian (China) during November 29 to December 1, 2003. Paper
for the proceedings and the proceedings to appear as a special issue of High 
Energy Phys. \& Nucl. Phys. (China).}
\vskip 0.15cm
Physical Research Laboratory, Ahmedabad 380 009, India		     
\end{center}		     
\vskip 0.25cm
\small{
{\bf Abstract:} Random one plus two-body hamiltonians invariant with respect
to $O(\cn_1) \oplus O(\cn_2)$ symmetry in the group-subgroup chains $U({\cal
N}) \supset U({\cal N}_1) \oplus U({\cal N}_2) \supset O({\cal N}_1) \oplus
O({\cal N}_2)$ and  $U({\cal N}) \supset O({\cal N}) \supset O({\cal N}_1)
\oplus O({\cal N}_2)$ chains of a variety of interacting boson models are
used to investigate the probability of occurrence of a given $(\omega_1
\omega_2)$ irreducible representation (irrep)  to be the ground state in
even-even nuclei; $\l[\omega_1\r]$ and $\l[\omega_2\r]$ are symmetric irreps
of $O(\cn_1)$ and $O(\cn_2)$ respectively.  Employing a 500 member random
matrix ensemble for $N$ boson systems (with $N=10-25$), it is found that for
$\cn_1,\cn_2 \geq 3$ the $(\omega_1 \omega_2)=(00)$ irrep occurs with $\sim
50\%$ and $(\omega_1 \omega_2)=(N0)$ and $(0N)$ irreps each with $\sim 25\%$ 
probability. Similarly, for $\cn_1 \geq 3, \cn_2=1$, for even $N$ the
$\omega_1=0$  occurs with $\sim 75\%$ and $\omega_1=N$ with $\sim 25\%$
probability and for odd $N$, $\omega_1=0$ occurs with $\sim 50\%$ and
$\omega_1=1,N$ each with $\sim 25\%$ probability. An extended Hartree-Bose
mean-field analysis is used to explain all these results.

\begin{center}
1. INTRODUCTION
\end{center}

Two-body random matrix ensembles (TBRE) defined over Hilbert spaces of 
various nuclear models led to the discovery that many of the regular
features observed in low-lying levels and near the yrast line in nuclei can
arise due to random interactions (with rotational symmetry) and this is
opposed to the  conventional ideas of using regular (or coherent)
interactions like pairing etc. in the nuclear hamiltonian. For the first
time this result is found by Bertsch et al \cite{Be-98} using the shell model 
who showed that with random interactions ground states in even-even nuclei
will be $0^+$ with very high probability and they also generate
odd-even staggering in binding energies, the seniority pairing gap etc.
Similarly, Bijker and Frank \cite{Bi-00} using the interacting boson model 
showed that random interactions generate vibrational and rotational
structures with high probability. These unexpected results gave rise to a
new field of research activity with random interactions in nuclei (they go
beyond the TBRE applications for smoothed (with respect to energy) state
densities, strength sums, transition matrix elements, information entropy in
wavefunctions etc. in nuclei and other finite quantum systems; see
\cite{Ko-01} and references therein). In particular: (i) Zelevinsky and
collaborators proposed the idea of geometric chaos for describing
regular features generated by TBRE's; (ii) Arima's group introduced a variety
of prescriptions, for predicting the probabilities, for simple systems such
as single $j$-shell for fermions, single $\ell$-shell for bosons and some of
their extensions; (iii) Bijker and Frank used a mean-field analysis with
projective coherent states for interacting boson systems. For details of
these studies we refer the readers to two recent reviews on this subject
\cite{Ze-03,Zh-03}. 

A very important aspect of TBRE's is that they admit group symmetries
\cite{Ko-02}. With $m$ particles (fermions or bosons) in $\cn$ single
particle states there is a $U(\cn)$ spectrum generating algebra (SGA). In
all the shell model/Interacting boson model studies reviewed in
\cite{Ze-03,Zh-03} one plus two-body hamiltonians that are rotational
scalars are considered, i.e. all the $O(3)$ scalars in $U(\cn) \supset
O(3)$ with one and two particle matrix elements of the one and two-body
parts respectively chosen to be random variables (in some studies the one
body part is dropped).  Immediately one sees that TBRE's can be extended to
scalars of various subgroups of $U(\cn)$, i.e. scalars of $G$ in $U(\cn)
\supset G \supset \ldots \supset O(3)$. The purpose of the present paper is
to consider such an extension to $O(\cn_1) \oplus O(\cn_2)$,
$\cn_1+\cn_2=\cn$ which appears in a very large class of interacting boson
models (IBM's) used in nuclear structure and address the question of with 
what probability a given $(\omega_1 \omega_2)$ irreducible representation
(irrep) will be the ground state in even-even nuclei; note that
$\l[\omega_1\r]$ and $\l[\omega_2\r]$ are symmetric irreps
of $O(\cn_1)$ and $O(\cn_2)$ respectively. In Section 2 given are the
random one plus two-body hamiltonians with $O(\cn_1) \oplus O(\cn_2)$
symmetry in interacting boson models. Section 3 gives the results of
numerical TBRE calculations and their understanding using an extended
Hartree-Bose mean-field analysis. Finally Section 4 gives conclusions and
future outlook.

\begin{center}
2. RANDOM INTERACTIONS WITH $O(\cn_1) \oplus O(\cn_2)$ SYMMETRY IN
IBM's
\end{center}

Large class of interacting boson models (IBM's) of nuclei admit $U({\cal
N}) \supset U({\cal N}_1) \oplus U({\cal N}_2) \supset O({\cal N}_1) \oplus
O({\cal N}_2)$ and  $U({\cal N}) \supset O({\cal N}) \supset O({\cal N}_1)
\oplus O({\cal N}_2)$ group-subgroup chains; $\cn=\cn_1+\cn_2$. Examples
(all for even-even nuclei) are: 
(i) $sp$IBM or nuclear vibron model \cite{Mod-1} with $U(4)$ SGA and 
$(\cn_1,\cn_2)=(3,1)$; (ii) $sd$IBM for quadrupole collective states
\cite{Mod-2} with $U(6)$ SGA and $(\cn_1,\cn_2)=(5,1)$; (iii) $spd$IBM for
GDR states  \cite{Mod-3} with $U(9)$ SGA and $(\cn_1,\cn_2)=(8,1)$,
$(6,3)$, $(5,4)$; (iv) $sdg$IBM for quadrupole plus hexadecupole states 
\cite{Mod-4} with $U(15)$ SGA and $(\cn_1,\cn_2)=(14,1)$, $(9,6)$,
$(10,5)$; (v) $sdpf$IBM for octupole states  \cite{Mod-5} with $U(16)$ SGA
and $(\cn_1,\cn_2)=(15,1)$, $(10,6)$ etc; (vi) $sdgpf$IBM \cite{Mod-6} with
$U(25)$ SGA and $(\cn_1,\cn_2)=(24,1)$, $(15,10)$ etc; (vii) $spp^\prime$IBM
or the $U(7)$ model for 3-body clusters in nuclei \cite{Mod-7} with $U(7)$
SGA and $(\cn_1,\cn_2)=(6,1)$, $(4,3)$; (viii) IBM-3 or the isospin ($T$)
invariant $sd$IBM (here the bosons carry $T=1$ degree of freedom)
\cite{Mod-8} with $U(18)$ SGA and $(\cn_1,\cn_2)=(15,3)$; (ix) IBM-4 or the
spin-isospin ($S,T$) invariant $sd$IBM (here the bosons carry $(ST)=(10)
\oplus (01)$ degree of freedom) \cite{Mod-9} with $U(36)$ SGA giving
examples with $(\cn_1,\cn_2)= (30,6)$, $(3,3)$,  $(18,18)$, $(15,15)$; 
(x) IBM-2 or proton-neutron IBM \cite{Mod-10} with $U(12)$ SGA  and
$(\cn_1,\cn_2)=(10,2)$. In this paper, for simplicity, we consider group
chains with $\cn_1 \geq 3, \cn_2=1$ and $\cn_1 \geq 3, \cn_2 \geq 3$,
i.e. $\cn_2=2$ situations (as in (x) above) are not  considered.

Group chains, for symmetric $U(\cn)$ irreps $\{N\}$  the irrep
labels for other group algebras in the chains and their reductions for the
$\cn_1 \geq 3$ and $\cn_2=1$ situation (hereafter called I) are,
$$
\l. \l| \barr{ccccccc} U(\cn) & \supset & U(\cn_1=\cn-1) & \supset & 
O(\cn_1) & \supset &
K \\ \{N\} & & \l\{n_1\r\} & & \l[\omega_1\r] & & \alpha \earr \r. \ran
$$
\be
n_1 = 0,\,1,\,2,\,\cdots,\,N,\;\;\;\; \omega_1 =n_1, n_1-2,\ldots,0\;
\mbox{or}\;1
\ee
and
$$
\l. \l| \barr{ccccccc} U(\cn) & \supset & O(\cn) & \supset & 
O(\cn_1) & \supset &
K \\ \{N\} & & \l[\omega\r] & & \l[\omega_1\r] & & \alpha \earr \r. \ran
$$
\be
\omega=N,N-2,\ldots,0\;\mbox{or}\;1,\;\;\;\;\omega_1=0,1,2,\ldots,\omega
\ee 
In (1,2), label(s) $\alpha$ for the irreps of $K$ need not be specified as 
the algebra $K$ do not play any role in the present work. Note that $U(\cn)
\supset U(\cn_1=N-1) \oplus U(\cn_2=1)$ and the $U(\cn_2=1)$ and its irreps
$\{n_2\}$, $n_2=N-n_1$ are not shown in (1). The general 
one plus two-body $O(\cn_1)$ ($O(\cn_2=1)$ will not exist)
scalar hamiltonian built out of
the Casimir operators of the group algebras in the chains (1,2) is given by
\be
\barr{rcl}
H^{I} & = & \frac{1}{N}\l[\alpha_1 C_1(U(\cn_1)) + \alpha_2 C_1(U(\cn_2=1)) 
\r] \\
& & + \frac{1}{N(N-1)}\l[ \alpha_3 C_2(U(\cn_1)) + \alpha_4 C_2(U(\cn_2=1))
+ \alpha_5 C_1(U(\cn_1))C_1(U(\cn_2=1)) \r. \\
& & + \l. \alpha_6 C_2(O(\cn)) + \alpha_7 C_2(O(\cn_1)) \r]
\earr
\ee
Similarly for the
$\cn_1 \geq 3$ and $\cn_2 \geq 3$ situation (hereafter called II),
the group chains, for symmetric $U(\cn)$ irreps $\{N\}$  the irrep
labels for other group algebras in the chains and their reductions are,
$$
\l. \l| 
\barr{ccccccccccc}
U(\cn) & \supset & U\l(\cn_1\r) & \oplus &  U\l(\cn_2\r) &
\supset & O\l(\cn_1\r) & \oplus & O\l(\cn_2\r) &  \supset & K \\
\{N\} & & \l\{n_1\r\} & & \l\{n_2\r\} & & \l[\omega_1\r] & &
\l[\omega_2\r] & & \alpha \earr
\r. \ran
$$
\be
\barr{l}
n_1 = 0,\,1,\,2,\,\cdots,\,N;\;\;\;n_2 \;= \; n-n_1 \\ 
\omega_1 =n_1, n_1-2,\ldots,0\;\mbox{or}\;1,
\;\;\;\omega_2 =n_2, n_2-2,\ldots,0\;\mbox{or}\;1
\earr
\ee
and
$$
\l. \l| 
\barr{ccccccccc}
U(\cn) & \supset & O(\cn) &
\supset & O\l(\cn_1\r) & \oplus & O\l(\cn_2\r) &  \supset & K \\
\{N\} & & \l[\omega\r] & & \l[\omega_1\r] & &
\l[\omega_2\r] & & \alpha \earr
\r. \ran
$$
\be
\omega=N,N-2,\ldots,0\;\mbox{or}\;1,\;\; \omega_1 + \omega_2 = \omega,
\omega-2, \ldots, 0\;\mbox{or}\;1
\ee
The general  one plus two-body $O(\cn_1) \oplus O(\cn_2)$ scalar hamiltonian
built out of the Casimir operators of the group algebras in the chains (4,5)
is given by
\be
\barr{rcl}
H^{II} & = & \frac{1}{N}\l[\alpha_1 C_1(U(\cn_1)) + \alpha_2 C_1(U(\cn_2)) \r]
\\
& & + \frac{1}{N(N-1)}\l[ \alpha_3 C_2(U(\cn_1)) + \alpha_4 C_2(U(\cn_2))
+ \alpha_5 C_1(U(\cn_1))C_1(U(\cn_2)) \r. \\
& & + \l. \alpha_6 C_2(O(\cn)) + \alpha_7 C_2(O(\cn_1)) + \alpha_8 
C_2(O(\cn_2)) \r]
\earr
\ee
In (3,6), $C_1$ and $C_2$ are linear and quadratic Casimir invariants and
their matrix elements for example are $\lan C_1(U(\cn_1)) \ran^{\{n_1\}} =
n_1$,  $\lan C_2(U(\cn_1)) \ran^{\{n_1\}} = n_1(n_1 + \cn_1 -1)$ and $\lan
C_2(O(\cn_1)) \ran^{\l[\omega_1\r]} = \omega_1(\omega_1 + \cn_1 -2)$. Given
$N$ bosons, in the $\l|N n_1 \omega_1\ran$ basis for I and $\l|N (n_1 n_2)
(\omega_1 \omega_2) \ran$ basis for II, the many boson H matrix for $H^I$ and
$H^{II}$ respectively will be always tridiagonal; the $\alpha_6$ terms in
Eqs. (3,6) generate off-diagonal matrix elements. In particular I is a
generalization of the $sp$IBM analyzed before \cite{Ku-00,Bi-01} while II is
completely new. For each allowed $\omega_1$ in I and $(\omega_1 \omega_2)$ in
II the $H$ matrices are constructed using the transformation brackets, given
in \cite{Ko-97}, between the chains (1) and (2) for I and (4) and (5)  for
II. The matrices are diagonalized, for each member of a 500 member TBRE, for
boson numbers $N=10-25$. Thus, in the calculations the parameters in Eqs.
(3,6) are chosen to be independent Gaussian variables with zero mean and unit
variance and 500 samples of the same are considered. Now we will discus the
results.

\begin{center}
3. RESULTS FOR $\cn_1 \geq 3, \cn_2=1$ AND $\cn_1 \geq 3, \cn_2 \geq 3$ 
SYSTEMS AND THEIR MEAN-FIELD ANALYSIS 
\end{center}

\begin{flushleft}
{\it 3.1 Results of Numerical Calculations}
\end{flushleft}

Figures 1a and 1b give the probabilities, in the situation $\cn_1 \geq 3$ and
$\cn_2=1$, i.e. for I as $\cn_1$ is varied. In Fig. 1a shown are the results 
for $\omega_1=0,N$ for even boson number $N$ ($N=10$) and in Fig. 1b for
$\omega_1=0,1,N$ for odd $N$ ($N=15$) to be ground states. Fig. 1c shows the
same  results but as a function of the boson number $N$ for $\cn_1=14$. It
should be noted that the probabilities are negligiblly small for the
$\l[\omega_1\r]$ irreps not shown in the figures.  In general for even $N$,
$\omega_1=0$ is ground state irrep with $\sim 65-74$\% and $\omega_1=N$ with
$\sim 25-32$\% probability. Similarly for odd $N$,  $\omega_1=0$ is ground
state irrep with $\sim 46-50$\%, $\omega_1=1$ with  $\sim 18-24$\%  and
$\omega_1=N$ with $\sim 24-30$\% probability. They reproduce the $sp$IBM
results known before \cite{Ku-00,Bi-01,Zh-03} and provide a test of the
present calculations. They are also close to the $sd$IBM results known before
(although in these studies $K=O(3)$ is chosen and $H^I$ to be a $O(3)$
scalar, note that $\omega_1=1$  gives $L=1$ in $sp$IBM and $L=2$ in $sd$IBM)
\cite{Bi-01,Bi-03,Zh-03}. 

\vskip -12in
\begin{center}
\epsfxsize 5in
\epsfysize 6in
\epsfbox{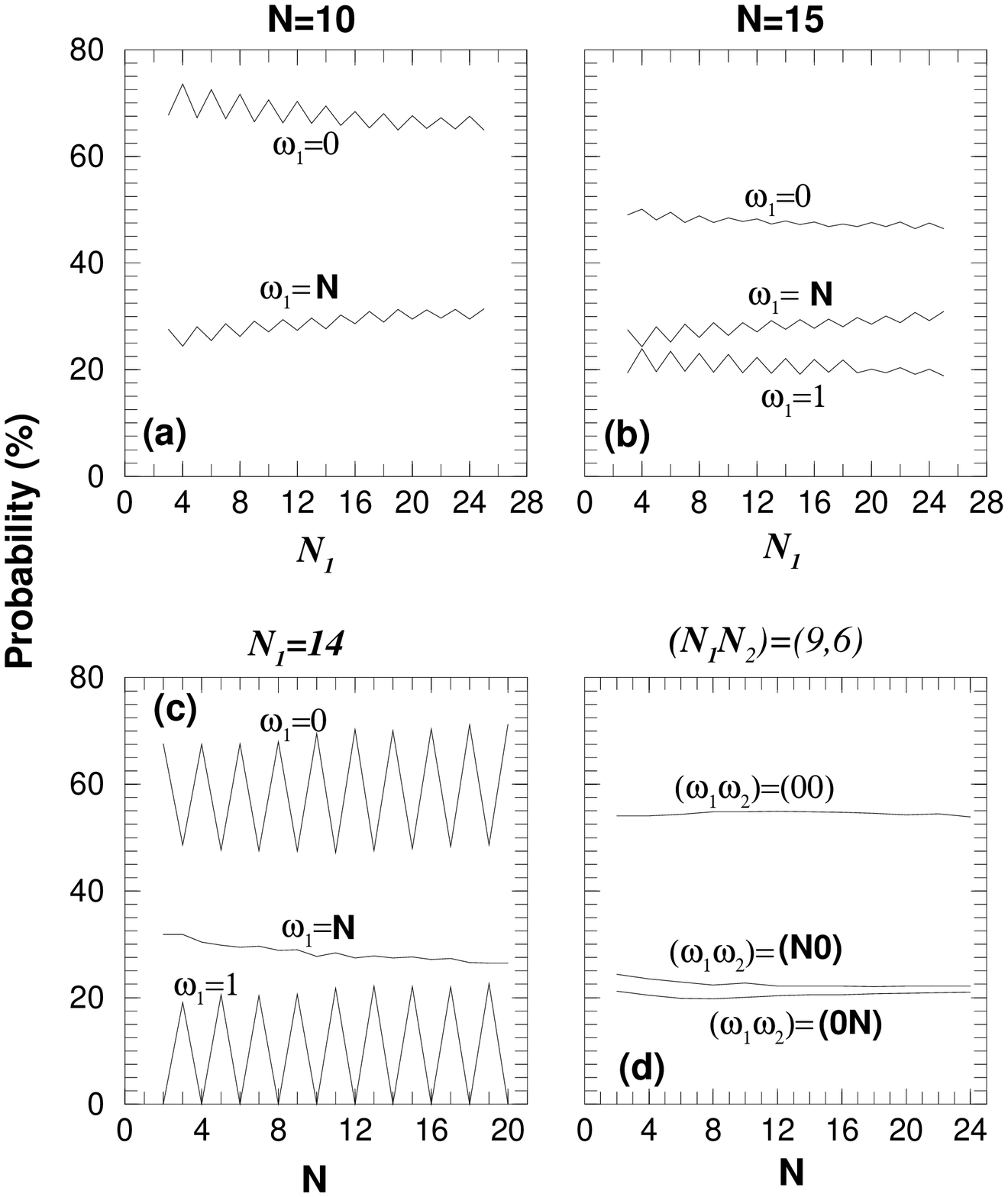} 
\end{center}
\vskip -3cm
\noindent {\bf Fig. 1.} {\footnotesize{
Probabilities for various group irreps to be ground
states. (a) and (b) give the variation as a function of $\cn_1$ for fixed
boson number and (c) and (d) as a function of the boson number $N$. (a), (b) 
and (c) are for I and (d) is for II. See text for details. }}

In the situation $\cn_1 \geq 3$, $\cn_2 \geq 3$ (i.e. II), Figure 1d (for
various $N$ values with $(\cn_1, \cn_2)=(9,6)$) and Table 1 (for various 
$(\cn_1, \cn_2)$ with $N=10,20$) give the results for the irreps $(\omega_1
\omega_2)=(00), (N0), (0N)$. Here only even $N$ is considered.  It is seen
that in general $(\omega_1 \omega_2)=(00)$ is ground state with $\sim
50-55$\% and $(\omega_1 \omega_2)=(0N)$ and $(N0)$ each with $\sim 20-24$\%
probability. The probabilities for other $(\omega_1 \omega_2)$ irreps to be
ground states is negligiblly small. Before giving a mean-field analysis of
these results some remarks are in order. In $sdg$IBM with $(\cn_1,
\cn_2)=(9,6)$, for $\gamma$-soft nuclei the $(\omega_1 \omega_2)=(0N)$ is
expected to be the ground state but, as seen from Fig. 1d, random
interactions give this irrep only with $\sim 20$\% probability. Therefore
random interactions are not good for $sdg$IBM for $(\cn_1, \cn_2)=(9,6)$.
However for $(\cn_1, \cn_2)= (14,1)$ (this system is used recently in phase
transition studies \cite{Pi-03}), the $U(14)$ irrep $\l[0\r]$ occurs, for
even $N$, as ground state with $\sim 70$\% and thus random interactions may
be useful here. Another example is, in IBM-4 with $(\cn_1, \cn_2)=(3,3)$
the $(\omega_1 \omega_2)=(ST)=(00)$ irrep is ground state with $\sim 50$\%
probability. However in real nuclei this irrep is expected to be the ground
state and therefore in IBM-4 one can use random interactions but a regular
part enhancing the probability for $(ST)=(00)$ should be added. 

\begin{flushleft}
{\it 3.2 Mean-field analysis}
\end{flushleft}

Bijker and Frank \cite{Bi-01,Bi-03} carried out a mean-field analysis of $sp$
and $sd$ IBM's to give a quantitative understanding of the probabilities,
with random interactions in these models, for a given $L$ to be the ground 
state. We will follow this approach with suitable extensions to describe the
results found in Fig. 1 and Table. 1. We begin with I, i.e. $\cn_1 \geq 3$,
$\cn_2=1$. The $\cn_2=1$ gives $s$ bosons (with angular momentum $\ell=0$).
Similarly $\cn_1 \geq 3$ gives bosons carrying $\cn_1$ degrees of freedom and
they can be thought of as bosons with $\ell=\ell_1, \ell_2, \ldots \ell_{r}$,
$\sum^{r}_{i=1} (2\ell_i +1) = \cn_1$. Just as in \cite{Bi-03}, a one
parameter H is considered (with $\alpha_1$ and $\alpha_6$ terms in (3)),
\be
H=\frac{1}{N}\,\cos \chi \,{\hat n}_1 + \frac{1}{N(N-1)}\,\sin \chi\; S_+
S_-\;,\;\;S_+ = s^\dagger s^\dagger - \sum_{i=1}^{r} b^\dagger_{\ell_i} \cdot   
b^\dagger_{\ell_i}\;,\;\;S_- = (S_+)^\dagger
\ee
In (7) $-\pi/2 < \chi \leq 3\pi/2$ so that all attractive and repulsive
interactions are included. The ground state shapes and hence the ground state
$\omega_1$ are determined by minimizing the energy functional for the axially
symmetric coherent state (CS),
\be
\l|N\;\alpha\ran = \frac{1}{\sqrt{N!}} \l(\cos \alpha\,s^\dagger +
\sin \alpha \,x^\dagger_0\r)^N \,\l|0\ran\;,\;\; x^\dagger_0 = \frac{1}{
\sqrt{r}} \sum_{i=1}^{r} b^\dagger_{\ell_i,0}
\ee
In Eq. (8), $-\pi/2 < \alpha \leq \pi/2$ and it gives for example correctly 
the CS used for $sdg$IBM in the past \cite{Mod-4}. The energy functional
$E(\alpha)= \lan N\;\alpha \mid H \mid N\;\alpha \ran = \cos \chi\, \sin^2
\alpha + \frac{1}{4} \sin \chi\,\cos^2 2\alpha$. The minima of $E(\alpha)$
and the corresponding shape parameters $\alpha$ of  the CS divide into three
classes: (i) $\alpha=0$ for $-\pi/2 < \chi \leq \pi/4$, i.e. in a  $3\pi/4$
range; (ii) $\alpha=\pi/2$ for $3\pi/4 \leq \chi \leq 3\pi/2$, i.e. in a
$3\pi/4$ range; (iii) $\alpha$ such that $\cos 2\alpha = \cot \chi$ for
$\pi/4 \leq \chi \leq 3\pi/4$, i.e. in a $\pi/2$ range. It is seen from (8)
that $\alpha=0$ gives $s$ boson condensate and hence here $\omega_1=0$ and it
occurs with $(3\pi/4(2\pi))\times 100$\% = $37.5$\% probability. Similarly
$\alpha=\pi/2$ gives $x$ boson condensate and  here, apart from a constant
factor, $E= -\sin\chi\,\omega_1(\omega_1 + \cn_1-2)$ where $\omega_1=n_1,
n_1-2, \ldots, 0$ or 1. Then clearly, $\sin \chi$ positive (this happens
$\pi/4$ times) gives $\omega_1=N$ to be  lowest with $12.5$\% probability and
$\sin \chi$ negative (this happens $\pi/2$ times) gives $\omega_1=0$ to be
lowest for even $N$ and $\omega_1=1$ for odd  $N$ with $25$\% probability.
For $\cos 2\alpha = \cot \chi$ the condensate is deformed with both $s$ and
$x$ bosons. It is plausible to argue that the condensate here gives a band
with $L=\kappa \omega_1$, $\omega_1=0,1,2, \ldots, N$ and $\kappa=1$ for
$sp$IBM, $\kappa=2$ for $sd$IBM, $\kappa=4$ for $sdg$IBM etc. The moment of
inertia (${\cal I}$) of these bands should follow from $O(\cn_1)$ cranking (a
method for this may be possible via the results in \cite{Le-87}). As yet
there is no theory for this and therefore we assume that the $O(3)$ cranking
formula given in  \cite{Bi-01,Bi-03} is valid here to within a constant. Then
(with ${\cal I} = (\sin \chi - \cos \chi)/\sin \chi \cos \chi$) it is easily
seen that for $\cos 2\alpha = \cot \chi$, $\omega_1=0$ is lowest with
$12.5$\% probability and $\omega_1=N$ is lowest with $12.5$\% probability.
Combining all these results will give for I for the ground state
probabilities: (i) $\omega_1=0$ with $75$\% and $\omega_1=N$ with  $25$\% for
even $N$; (ii) $\omega_1=0$ with $50$\%, $\omega_1=1$ with $25$\%  and
$\omega_1=N$ with $25$\% for odd $N$. They give a good description of the
results in Figs. 1a, 1b and 1c.

\begin{center}
{\bf Table 1.} Probabilities (in percentage) for $(\omega_1 \omega_2)$ to be 
ground state irrep
\vskip 0.2cm
{\scriptsize{
\begin{tabular}{ccccccc}
\hline
Model & $\;\;\;\cn_1\;\;\;$ & $\cn_2\;\;\;$ & $N\;\;\;$ & $(\omega_1
\omega_2)=(00)$ & $(\omega_1 \omega_2)=(N0)$ & $(\omega_1 \omega_2)=(0N)$ \\
\hline
$U(7)$ & 4 & 3 & 10 & 55.4 & 21.4 & 20.5 \\
& & & 20 & 54 & 21.2 & 20.6 \\
$spd$IBM & 6 & 3 & 10 & 55 & 22.4 & 19.8 \\
& & & 20 & 53.5 & 21.9 & 20.3 \\
$sdg$IBM & 10 & 5 & 10 & 55.3 & 22.9 & 19.2 \\
& & & 20 & 53.6 & 22.8 & 20.4 \\
$spdf$IBM & 10 & 6 & 10 & 49.3 & 24.6 & 21.9 \\
& & & 20 & 49.3 & 23.8 & 22 \\
$sdgpf$IBM & 15 & 10 & 10 & 53.8 & 22.9 & 20.3 \\
& & & 20 & 54.4 & 22.4 & 20.5 \\
IBM-3 & 15 & 3 & 10 & 49 & 27.1 & 19.8 \\
& & & 20 & 49 & 25.6 & 20.7 \\
IBM-4 & 3 & 3 & 10 & 49.2 & 22.8 & 22.8 \\
& & & 20 & 48.8 & 23.2 & 23.2 \\
& 30 & 6 & 10 & 50.1 & 28.6 & 18.6 \\
& 18 & 18 & 10 & 50 & 23.5 & 23.5 \\
& 15 & 15 & 10 & 49.6 & 23.6 & 23.6 \\
\hline
\end{tabular}
}}
\end{center}

Now we will consider the mean-field analysis for II, i.e. for  $\cn_1, \cn_2
\geq 3$ and the discussion will be restricted to even $N$. Just as the
$x^\dagger$ operator in Eq. (8), let us introduce $y^\dagger$ and $z^\dagger$
operators, $y^\dagger_0 = \frac{1}{ \sqrt{p} } \sum_{ i= 1}^{p}\, b^\dagger_{
\ell_i,0 }$,  $\sum^{p}_{i=1} (2\ell_i +1) = \cn_1$ and $z^\dagger_0 =
\frac{1}{ \sqrt{q} } \sum_{ j= 1}^{q}\, b^\dagger_{ \ell^\prime_j,0 }$, 
$\sum^{q}_{j=1} (2\ell^\prime_j +1) = \cn_2$. Then the hamiltonian, CS and
$E(\alpha)$ are,
\be
\barr{l}
H=\frac{1}{N}\,\cos \chi \,{\hat n}_2 + \frac{1}{N(N-1)}\,\sin \chi\; S_+
S_-\;, \nn8
S_+ = S_+(1) - S_+(2) = \dis\sum_{i=1}^{p} b^\dagger_{\ell_i} \cdot   
b^\dagger_{\ell_i} - \sum_{j=1}^{q} b^\dagger_{\ell^\prime_j} \cdot   
b^\dagger_{\ell^\prime_j}\;,\;\;S_- = (S_+)^\dagger \nn8
\l|N\;\alpha\ran = \frac{1}{\sqrt{N!}} \l(\cos \alpha\,y^\dagger_0 +
\sin \alpha \,z^\dagger_0\r)^N \,\l|0\ran \nn8
E(\alpha) = \cos \chi\,\sin^2\alpha + \frac{1}{4} \sin \chi\,\cos^2 2\alpha
\earr
\ee 
The $H$ in (9) is same as $H^{II}$ defined in Eq. (6) but only with the
$\alpha_2$ and $\alpha_6$ terms. The equilibrium shapes correspond to
$\alpha=0$, $\alpha=\pi/2$ and $\cos 2\alpha = \cot \chi$ with the range of
$\chi's$ just as before. The $\alpha=0$ gives $y$-boson condensate with
energy $E(\alpha=0) \propto  - \sin \chi\,\omega_1 (\omega_1 + \cn_1-2)$.
Then the ground state irreps are  $(\omega_1 \omega_2)=(00)$ with $25$\% and
$(\omega_1 \omega_2)=(N0)$ with $12.5$\% probability. Similarly
$\alpha=\pi/2$ gives $z$-boson condensate with energy $E(\alpha=\pi/2)
\propto - \sin \chi\,\omega_2 (\omega_2 + \cn_2-2)$ and then the ground state
irreps are $(\omega_1 \omega_2)=(00)$ with $25$\% and $(\omega_1
\omega_2)=(0N)$  with $12.5$\% probability. In the situation $\cos 2\alpha =
\cot \chi$, cranking has to be done with respect to both $O(\cn_1)$ and
$O(\cn_2)$. Evaluating moment of inertias as before gives $E$ to be, to
within a constant,
\be
E= \dis\frac{ \omega_1 (\omega_1 + \cn_1-2)}{A_+} +  \dis\frac{\omega_2
(\omega_2 + \cn_2-2)}{A_-}\;, \;\;A_\pm = \mp \dis\frac{(\sin \chi \pm
\cos \chi)}{\cos \chi\, \sin \chi}
\ee
With $\pi/4 \leq \chi \leq 3\pi/4$ here, it is seen that $A_+$ is +ve and
$A_-$ is $-$ve for $\pi/4 \leq \chi \leq \pi/2$ and $A_+$ is $-$ve and $A_-$
is +ve for $\pi/2 \leq \chi \leq 3\pi/4$. Therefore, here $(N0)$ and $(0N)$
irreps will be ground states each with $12.5$\% probability. Combining all
the results give for II, $(\omega_1 \omega_2) = (00)$, $(N0)$ and $(0N)$
irreps to be ground states with $50$\%, $25$\% and $25$\% probability. These
numbers clearly describe the results in Table 1 and Fig. 1d.  

In summary, the mean-field approach of \cite{Bi-01,Bi-03} with proper
extensions gives a good understanding of the results in Fig. 1 and Table 1
although all the results are obtained using a constant probability for $\chi$
in Eqs. (7,9).  Extension of the analysis to odd $N$ for II (here the lowest 
$(\omega_1 \omega_2)$ are $(10)$ and $(01)$) and also the calculations for
$\cn_2=2$ will be given elsewhere.

\begin{center}
4. CONCLUSIONS AND FUTURE OUTLOOK
\end{center}

In this paper for the first time TBRE's preserving irreps of group symmetries
(other than $O(3)$), for boson systems, are introduced and showed that the
$0^+$ dominance observed  in ground states extends to group irreps. An
extended mean-field analysis is shown to give good description of the
numerical results obtained for a variety of interacting boson models. The
mean-field analysis in Section 3 is restricted to the simple mixing
hamiltonians given by Eqs. (7,9) and  in a future paper this will be extended
to the full hamiltonians given by Eqs. (3,6). Similarly, in a future
publication we will consider $O(\cn)$, $\cn > 3$ cranking so that the
application of $O(3)$ cranking to I and II in Section 3 can be validated. At
present the justification for using $O(3)$ cranking comes from the good
agreement between the mean-field results and those in Fig. 1 and Table 1. It
should be added that Kusnezov's analysis for $sp$IBM \cite{Ku-00}, based on
random polynomials, can be applied to $H^I$ and $H^{II}$ as the matrices here
are tridiagonal and this will be done elsewhere. Finally, it will be
interesting to extend the present work to other general classes of
group-subgroup chains in IBM's (see \cite{Mod-8,Mod-9} for examples) and also
to group chains for fermion systems (as they appear for example in the shell
model \cite{Ko-03}). It is plausible that the results of these extensions
will give deeper understanding of geometric chaos and regularities generated
by random interactions.

\vskip 0.5cm

{\it  A. Arima's talk in  `Symmetries in Science XIII' held at Bregenz in
July 2003 has provided the basis for initiating the work presented in this
paper.  The author thanks N. Yoshinaga for discussions in the Bregenz meeting
and Y.M. Zhao for very useful correspondence. This article is dedicated to
Prof. J.P. Draayer on his 60$^{th}$ birthday.}

}
\ed